\documentclass[proceedings, preprint]{rmaa}
\pdfoutput=1
\usepackage{paralist}

\usepackage{psfrag,color}

\usepackage{graphics,graphicx,microtype}
\usepackage[varg]{txfonts}
\SetYear{2013}
\SetConfTitle{Magnetic Fields in the Universe IV}

\title{Probing Magnetic Fields using the Giant Metrewave Radio Telescope} 

\author{
  J. S. Farnes,\altaffilmark{1,2} 
  D. A. Green,\altaffilmark{1}
  and N. G. Kantharia\altaffilmark{3}}

\altaffiltext{1}{Astrophysics Group, Cavendish Laboratory, University of Cambridge, 19 J.~J.~Thomson
Avenue, Cambridge, CB3 0HE, UK.}

\altaffiltext{2}{Sydney Institute for Astronomy, School of Physics, University of Sydney, NSW 2006, Australia (jamie.farnes@sydney.edu.au).}

\altaffiltext{3}{National Centre for Radio Astrophysics, Tata Institute of Fundamental Research, Ganeshkhind, Pune -- 411007, India.}

\shortauthor{Farnes, Green \& Kantharia}
\shorttitle{Probing Magnetic Fields using the GMRT}

\listofauthors{J. S. Farnes, D. A. Green, \& N. G. Kantharia}
\indexauthor{Farnes, J. S.}
\indexauthor{Green, D. A.}
\indexauthor{Kantharia, N. G.}

\abstract{We present the first spectropolarimetric radio observations that apply Rotation Measure (RM) Synthesis to interferometric data from the Giant Metrewave Radio Telescope (GMRT) at 610~MHz. Spectropolarimetry requires measurement of a large number of instrumental systematics so that their effects can be calibrated -- a technical problem that is currently being faced by the upcoming SKA pathfinders. Our fully-calibrated data allow for the peak Faraday depth and polarisation fraction to be measured for sub-mJy compact sources in the field of M51 at 610~MHz. The diffuse extended emission in the interacting galaxy pair is shown to be depolarised below the sensitivity limit. A survey of linear polarisation with the GMRT is now feasible and could be used to place constraints on the prevailing depolarisation mechanisms at low frequencies -- improving polarised source count estimates and constraining the RM-grid observable with next generation facilities such as the SKA.}

\resumen{Presentamos las primeras observaciones espectropolarim\'etricas que
aplican s\'\i ntesis de medidas de rotaci\'on a datos
interferom\'etricos obtenidos con el Giant Metrewave Radio Telescope
(GMRT) a 610~MHz.  
La espectropolarimetr\'\i a requiere la medici\'on
de un gran n\'umero de efectos sist\'ematicos instrumentales para
lograr calibrar sus efectos -- un reto t\'ecnico al que se enfrentan
los instrumentos precursores del SKA. Nuestros datos completamente
calibrados permiten medir el pico de la profundidad de Faraday a nivel
de sub-mJy en fuentes compactas en el campo de M51 a 610~MHz. 
Se muestra que la radiaci\'on difusa extendida en el par de galaxias
interactuantes es depolarizada por debajo del l\'\i mite de
sensibilidad. Un censo de polarizaci\'on lineal con el GMRT es ahora
posible, y podr\'\i a ser utilizado para delimitar los mecanismos de
depolarizaci\'on a frecuencias bajas -- mejorando as\'{\i} las
estimaciones y l\'\i mites de en las observaciones de medidas de
rotaci\'on de las nuevas generaciones de instrumenros, como el SKA.}

\addkeyword{Galaxies: Individual (NGC~5194, NGC~5195)}
\addkeyword{Magnetic fields}
\addkeyword{Polarization}
\addkeyword{Techniques: Interferometric}
\addkeyword{Techniques: Polarimetric}

\begin{document}
% Typeset article header
\maketitle

\section{Introduction}
\label{sec:intro}
Polarised radio emission is fundamentally related to the presence of magnetic fields, and observations of polarisation are arguably the best way of directly studying quasi-regular fields \citep[e.g.][]{2013arXiv1302.0889B}. A key issue for polarimetry of astrophysical sources is removal of effects that alter the properties of the radiation as it propagates across the Universe and through the various components of a radio telescope \citep[e.g.][]{HamakerREF}.

Radio polarisation observations require extensive analysis in order to prepare a facility for the complicated nature of such measurements. Such commissioning efforts allow for the processing of data in which all four cross-correlation products have been recorded i.e.\ \(RR\), \(LL\), \(RL\), \(LR\) \citep[e.g.][]{mythesis}. The ``full-polarisation'' mode of the Giant Metrewave Radio Telescope (GMRT) has recently become available for such observations. Imaging of GMRT polarisation data has been previously attempted in the 610~MHz band using the bandwidth-averaged Stokes $Q$/$U$ (Joshi \& Chengalur 2010). Nevertheless, as with all new instrumental modes of operation, the facility requires efforts to understand the various intrinsic limitations to such observations, such as the instrumental polarisation, instrumental time-stability, quality of $uv$-data calibration, ionospheric Faraday rotation, polarisation angle corrections across the observing bandwidth, and the wide-field response. Wide-field polarimetric effects and their removal are a particularly important consideration, as many of the upcoming SKA pathfinders will use vast fields of view to explore magnetic fields.

ASKAP will observe at frequencies between 700~MHz--1.8~GHz \citep[e.g.][]{2008ExA....22..151J}, GALFACTS between 1.2--1.5~GHz \citep[e.g.][]{2010ASPC..438..402T}, LOFAR at $<$230~MHz \citep[e.g.][]{2012arXiv1203.2467A}, and the upgraded Jansky-VLA (JVLA) at $\gtrsim$1.0~GHz \citep[e.g.][]{2009IEEEP..97.1448P}. The GMRT itself is currently undergoing an upgrade that will further improve upon its science capabilities (Gupta, 2011). When the upgrade is complete, the intended nearly-seamless frequency coverage from 150~MHz to 1.5~GHz, with instantaneous bandwidths of 400~MHz, will therefore be centred within a frequency range with limited complementary observational data. Wide-field polarimetric surveys with the GMRT therefore have the potential to fill a crucial gap in our understandings of cosmic magnetism \citep[e.g.][]{2012A&A...543A.113B}.

Here we show that such observations are now possible at the GMRT at 610~MHz, and that wide-field spectropolarimetry can be carried out. We demonstrate the polarisation capabilities of the instrument by observing the nearby galaxy M51 (NGC~5194). This provides the opportunity to assess the quality of polarisation calibration, and to study the polarisation properties of the galaxy at relatively low frequencies. The galaxy and its linear polarisation properties have been well-studied at frequencies $\gtrsim$1.4~GHz (see e.g.\ Horellou et al.\ 1992; Fletcher et al.\ 2011 and references therein). The same techniques have also been applied to Southern Compact Group galaxies in order to look for evidence of interactions between the group members and the surrounding intergalactic medium, and to study the magnetic fields in radio galaxies (Farnes et al., in preparation).

\section[]{Observations and Data Reduction}

The GMRT observations were centred at 610~MHz with a 16~MHz bandwidth that was equally divided into 256 channels. The observations used the flux calibrators 3C48 and 3C286, and were taken in GMRT Cycle 17 in January 2010. The observation used the phase calibrator J1313+549, which was found to have a derived flux density of \(1.719\pm0.008\)~Jy. Unless otherwise stated, data reduction and calibration were carried out using standard tasks in the 31DEC10 AIPS package. 

\begin{figure}[t]
\centering
\hspace{-48pt} \resizebox{60mm}{!}{\includegraphics[clip=true,trim=0.6cm 1cm 12.5cm 0.6cm]{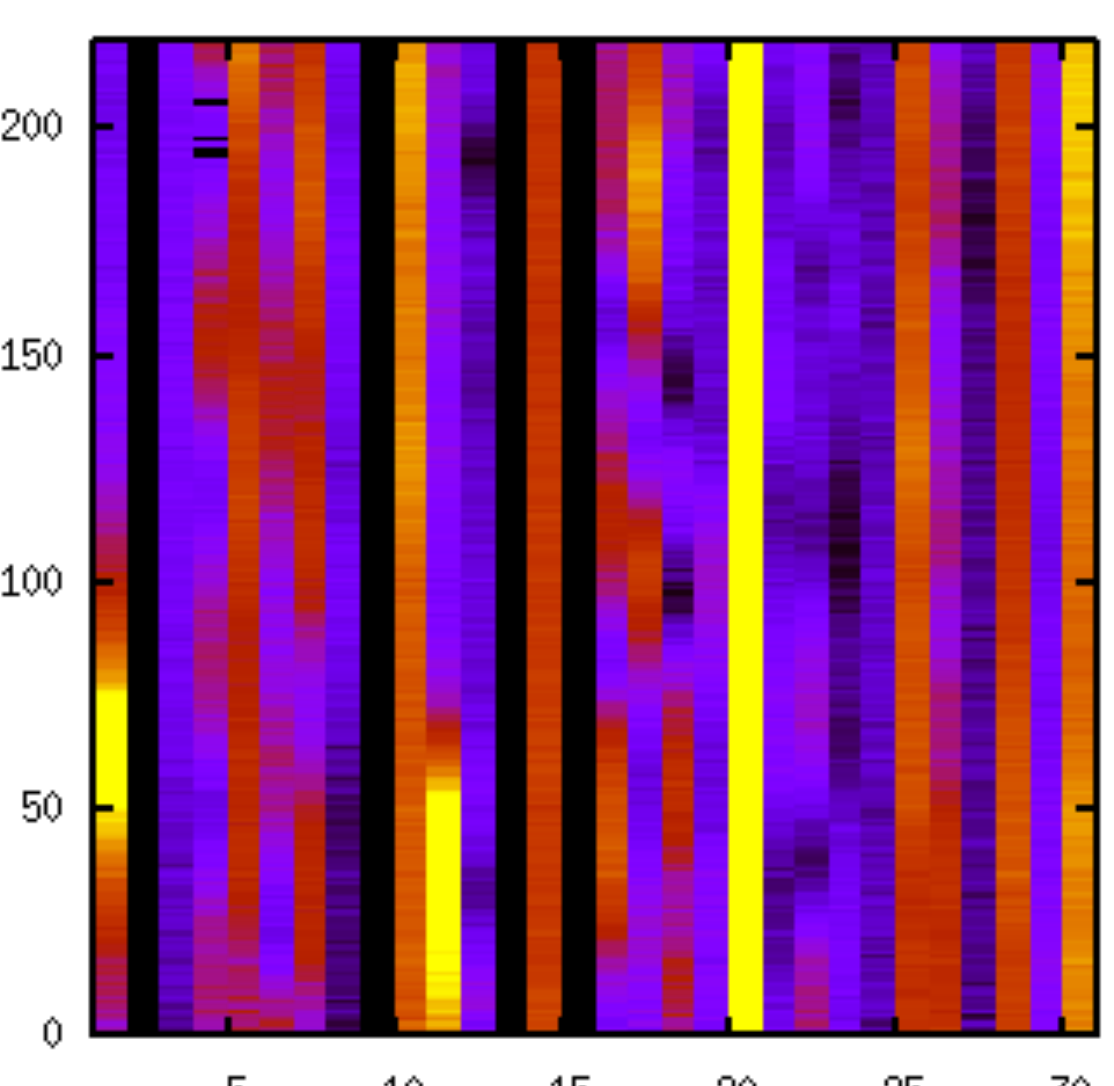}} \\
  \vspace{-5pt}
  \resizebox{66mm}{!}{\includegraphics[clip=true,trim=12cm 0cm 0cm 0.7cm]{M51leakages}}
  \vspace{0pt}
  \caption{\small{The leakage amplitude for the observation of M51, in both \(R\) (top) and \(L\) (bottom). The $x$-axis shows the antenna number, the $y$-axis shows the channel number across the observing bandwidth, and the pseduo-colour scale shows the leakage amplitude. Antennas coloured black across the entire band are flagged.}}
  \label{leakages}
  \vspace{0pt}
\end{figure}

The antenna-based instrumental polarisation (`leakage') was found to be highly frequency-dependent \citep[also see][]{2009ASPC..407...12T}, so the leakages must therefore be calculated for each individual spectral channel. The typical leakage amplitude is $\approx5$--$10$\%, with a few antennas having leakages of up to \(\sim40\)\%, as shown in Figure \ref{leakages}. Separation of instrumental and source contributions was achieved through repeated observation of the phase calibrator over a large range in parallactic angle. This was done with the AIPS task \textsc{pcal} using a linearised model of the feeds (\textsc{soltype=`appr'}), as a full non-linearised model is not currently available in commonly used data reduction packages. A full discussion of the GMRT technical capabilities and data reduction process will be described elsewhere (Farnes et al., in preparation).

Following leakage calculation, checks on the calibrators 3C48 and 3C286 show that the residual instrumental polarisation is \(\le0.25\)\%. The electric vector polarisation angle (EVPA) was corrected in each spectral channel using 3C286, which was taken to have an RM of \(-1.2\)~rad~m\(^{-2}\) and intrinsic EVPA of 33$^{\circ}$.

Each channel was imaged in Stokes \(Q\)/\(U\)/\(V\) individually using multi-facet imaging -- breaking the sky up into 31 facets across the field of view. The resulting Stokes \(Q\) and \(U\) maps were combined into images of polarised intensity, \(P = \sqrt{Q^2 + U^2}\). Some channels at the edges of the band were discarded, resulting in 220 images of \(P\). These were initially averaged together into an image of the band-averaged polarised intensity -- which provides a useful diagnostic of the polarisation across the field of view (FOV).

The Stokes $Q$ and $U$ images were then used to form a three-dimensional datacube, so that the technique of RM Synthesis could be applied \citep{2005A&A...441.1217B}, using a code developed in Python. This code also implemented a form of RM-clean (Heald et al. 2009) to deconvolve the Faraday dispersion function (FDF) from the Rotation Measure Spread Function (RMSF). The RM Synthesis observational setup, which uses a 16~MHz bandwidth, with 62.5~kHz channel spacing at 610~MHz, has a maximum Faraday depth to which one is sensitive of $|\phi_{max}| \sim 35000$~rad~m$^{-2}$, and a FWHM of the RMSF of 321~rad~m$^{-2}$. No attempt was made to distinguish multiple Faraday components along a single line of sight. Bandwidth depolarisation will affect these data for a $|$RM$|\ge20000$~rad~m$^{-2}$. The maximum scale in Faraday space to which the data are sensitive is 14~rad~m$^{-2}$ -- the data are not sensitive to Faraday thick emission. Faraday depths (FDs) were retrieved using a least-squares Gaussian fit to the peak of the deconvolved FDF.

\begin{figure}
\centering
  \resizebox{82mm}{!}{\includegraphics[clip=true,trim=2cm 0.2cm 0cm 0cm]{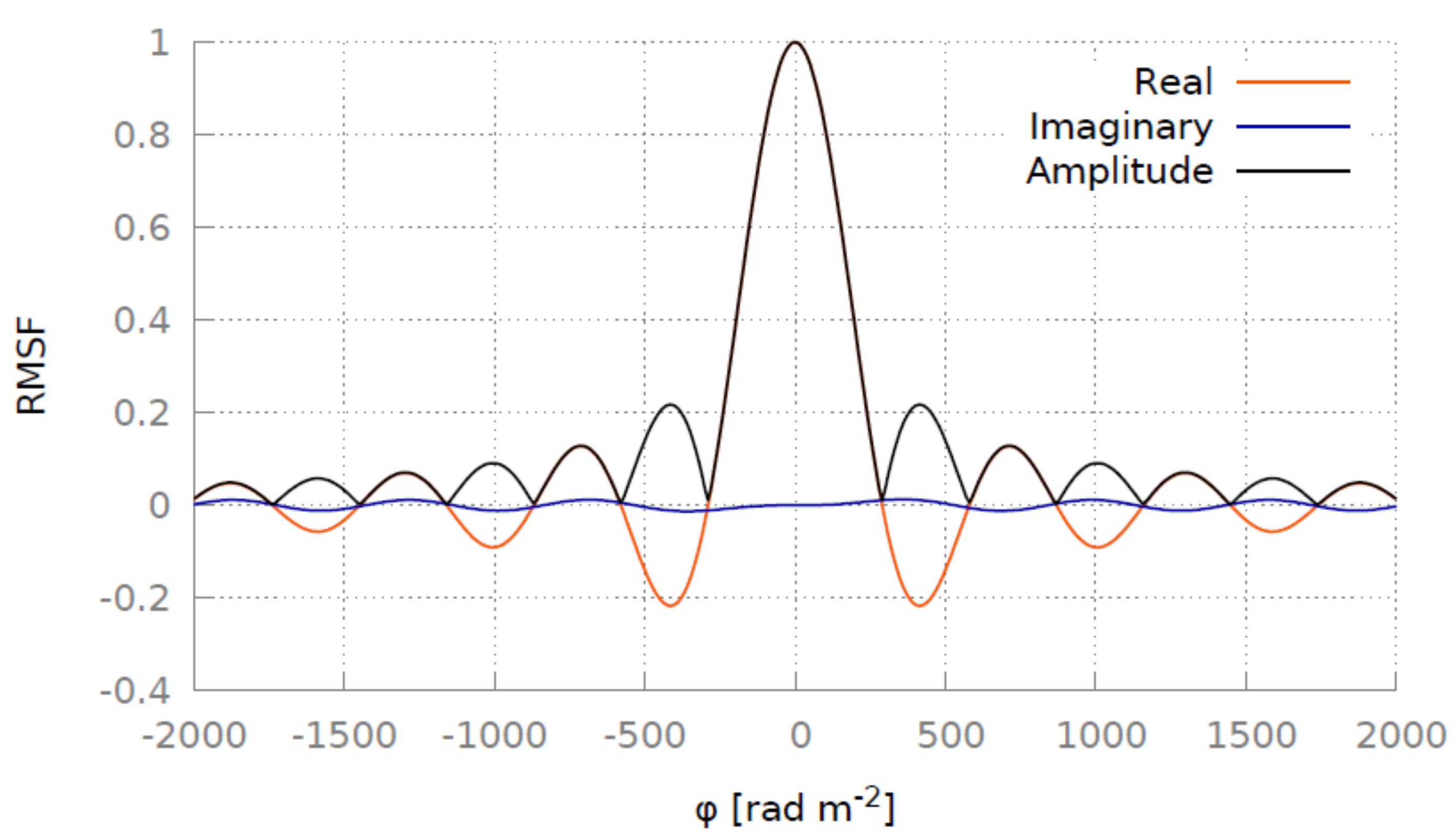}}
  \vspace{0pt}
  \caption[The RMSF for these observations]{\small{The rotation measure spread function (RMSF) for these GMRT observations. The real and imaginary components (\(Q+iU\)) and the amplitude (\(P = \sqrt{Q^2 + U^2}\)) are shown.}}
  \label{RMSF}
  \vspace{0pt}
\end{figure}

\section{Direction-dependent Instrumental Polarisation}
A limiting factor for wide-field polarimetric observations are the effects of beam `squint'/`squash' that result in polarimetric aberrations across the field of view. A classical model of beam squint/squash in prime-focus feeds suggests that beam squash can arise in \(Q\)/\(U\) due to two effects \citep[e.g.][]{2002ASPC.proc..278S,HeilesGBTMemo}: (i) the interaction of the linearly polarised vectors with the curvature of the reflecting surface, and (ii) the difference between the two polarisations in illumination of the primary surface (which occurs from pseudo-waveguide modes in a feed). Similarly, beam squint should arise in \(V\) due to the feed not pointing directly at the vertex of the paraboloid. There is no anticipation of beam squint in \(Q\)/\(U\) or beam squash in \(V\) \citep{HeilesGBTMemo}. The fundamental cause of direction-dependent instrumental linear polarisation is therefore the curved reflecting surface, which slightly changes the direction of an incident electric vector upon reflection. For on-axis sources, these polarimetric aberrations largely cancel out. For off-axis sources, the path length to the source from different parts of the reflecting surface are not all equal. These distortions increase with curvature \citep[see][for further details]{1996aspo.book.....T,2002ASPC.proc..278S}. 

A lot can be inferred about the GMRT's off-axis response from a conventional observation. The observation of M51 has \(\sim6\)~hours on source, with parallactic angle variation of M51 ranging from \(-86^{\circ}\) to \(+89^{\circ}\), with similar coverage for the nearby phase calibrator. This observation was used to investigate the effects of direction-dependent instrumental polarisation.

RM Synthesis was performed on the entire FOV surrounding M51, using images that were uncorrected for the Stokes \(I\) beam. The Faraday spectrum was processed between \(\pm2000\)~rad~m\(^{-2}\) and sampled every \(1\)~rad~m\(^{-2}\). Sources with peak brightness above a \(5\sigma\) threshold in Stokes \(I\) were manually identified. From the brightest pixel in Stokes \(I\), those sources with a peak brightness \(\ge8\sigma\) in the amplitude of their Faraday spectrum\footnote{The detection statistics differ in Faraday space. A \(7\sigma\) detection threshold in \(\phi\)-space has an equivalent false-detection rate to the \(\approx2.5\sigma\) level in total intensity. At \(8\sigma\), the false-detection rate is less than 1 in 33,000 \citep{2011arXiv1106.5362G}.} were considered polarised. These sources had their peak brightness in \(P\) calculated by fitting a Gaussian to data points surrounding the peak. The calculated peaks in \(P\) were corrected for the effects of Rician bias using \(\sigma\) from the cleaned \(Q\)/\(U\) \(\phi\)-cubes \citep[see][and references therein]{2011arXiv1106.5362G}. The corrected values were used to calculate the fractional polarisation, \(\Pi = P_{0}/I\). This procedure can underestimate the polarisation fraction, as sources may have polarisation structure that is offset from the peak in Stokes \(I\). Nevertheless, the calculated values still provide a good estimate of the polarised intensity resulting from off-axis effects.

Sources with a peak brightness less than \(8\sigma\) in the Faraday spectrum were considered unpolarised down to the sensitivity limit. The peak in the \(P\) image created from band-averaged \(Q\) and \(U\) was used to place an upper limit on the polarisation fraction of unpolarised sources. This procedure tends to overestimate the polarisation fraction, as the bias correction does not work well for low signal-to-noise levels \citep{1985A&A...142..100S}. Nevertheless, the calculated values serve as an upper limit on the polarisation fraction.

\begin{figure}[hbp]
\centering
    \resizebox{82mm}{!}{\includegraphics[height=82mm,clip=true,trim=2cm 0cm 0cm 0cm]{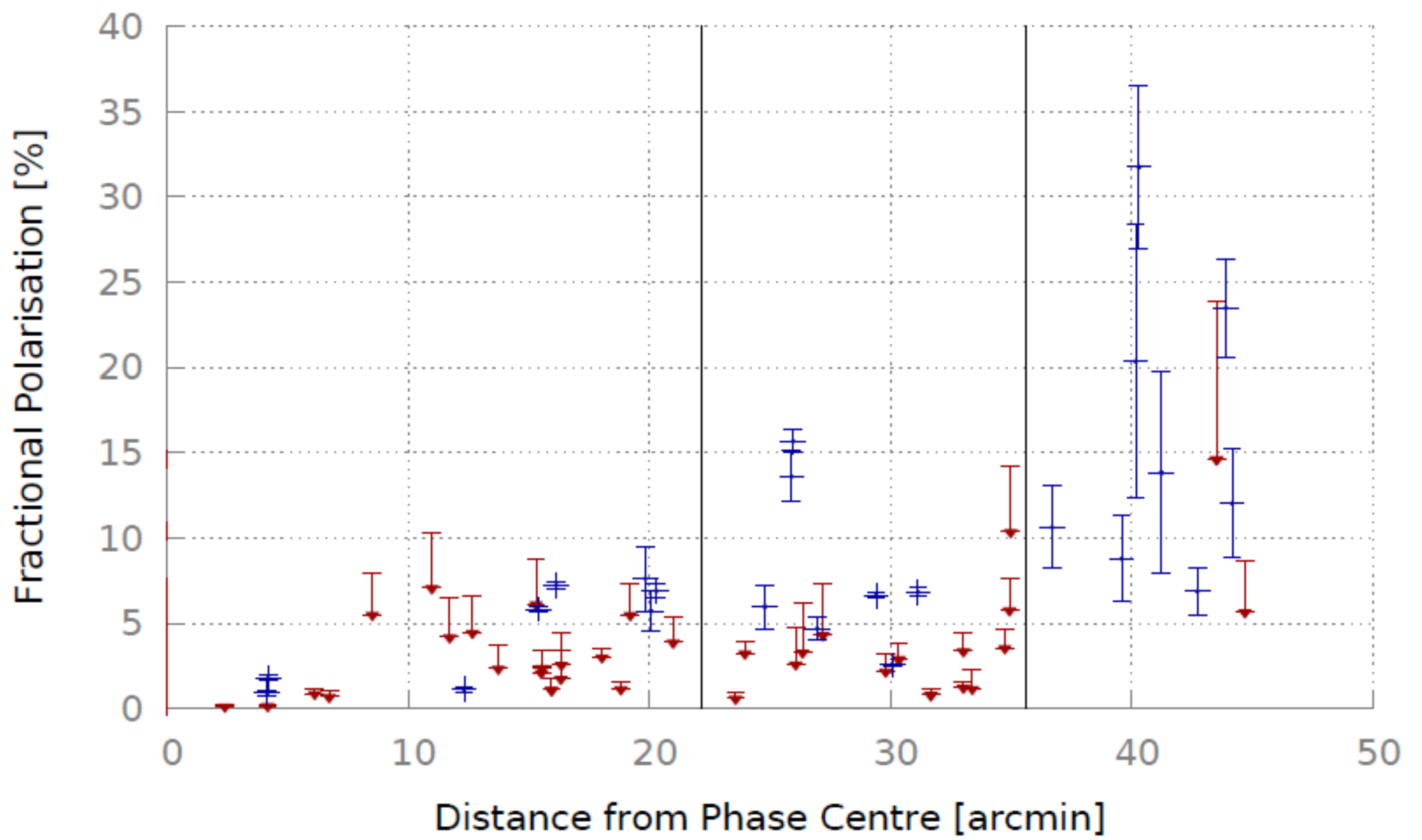}}
%  \vspace{-5pt} \\
  \caption[The fractional polarisation of sources versus distance from the phase-centre]{\small{The fractional polarisation, \(\Pi\), of sources surrounding M51 as a function of distance from the phase-centre. Sources detected above an \(8\sigma\) limit following RM Synthesis are shown in blue. Sources undetected in polarisation within the sensitivity limit (\(P < 8\sigma\)) have an upper bound shown in red. The half-power and \(10\%\) points of the Stokes \(I\) beam are indicated by the solid vertical lines.}}
  \label{chap6:PBefore}
\end{figure}

A plot of the fractional polarisations of both polarised and unpolarised sources as a function of increasing distance from the phase-centre is shown in Figure \ref{chap6:PBefore}. The plot provides a \emph{crude} estimate of the direction-dependent instrumental polarisation. It must be remembered that M51 - which is located near the phase-centre with an angular size covering several arcminutes - acts as a depolarising Faraday screen. Of the data beyond \(10^{\prime}\), there is no clearly discernible increase in \(\Pi\) out to the half-power point. Note that a statistical analysis would not be particularly useful, as the intrinsic and off-axis polarisation cannot be trivially separated. More rigourous constraints on the polarisation beam requires holography observations across the FOV -- see Farnes, (2012) for a detailed analysis. Between half-power to 10\% of the Stokes \(I\) beam, the direction-dependent response does appear to steadily increase. The interpretation is hindered by the two polarised sources with \(\Pi>10\)\% in this region of the beam, as it is not clear if these are intrinsically of higher polarisation or are instead affected by beam effects. Beyond the 10\% point of the beam, \(\Pi\) increases rapidly.

A direction-dependent response is apparent, with the spurious polarised intensity of sources visibly increasing as a function of distance from the phase-centre. A conservative `upper limit' on the effect of the polarisation beam at 610~MHz appears to be \(\le2.5\)\% at the half-power point, with a rapid increase beyond this to \(>20\)\% at the edge of the beam. Nevertheless, if the polarisation beam is oriented radially, it would be expected that this observation's large range in parallactic angle will have caused the direction-dependent instrumental effects to average down considerably. Therefore these `upper limits' truly provide a lower bound. 
%  \vspace{-10pt} 
% \vspace{-15pt} 

For the purpose of understanding the GMRT, it is also essential to check the frequency response of the off-axis polarisation. Although different instruments, the WSRT is known to have a frequency-dependent beam pattern with a period of \(\sim17\)~MHz \citep{2011AJ....141..191F}. The VLA also showed a strong change in off-axis characteristics as a function of IF \citep{CottonAIPSMemo}, while the beam polarisation of the JVLA is essentially frequency-independent across an observing band.

\begin{figure}[bhtp]
\centering
    \resizebox{82mm}{!}{\includegraphics[height=82mm,clip=true,trim=0cm 0cm 0cm 0cm]{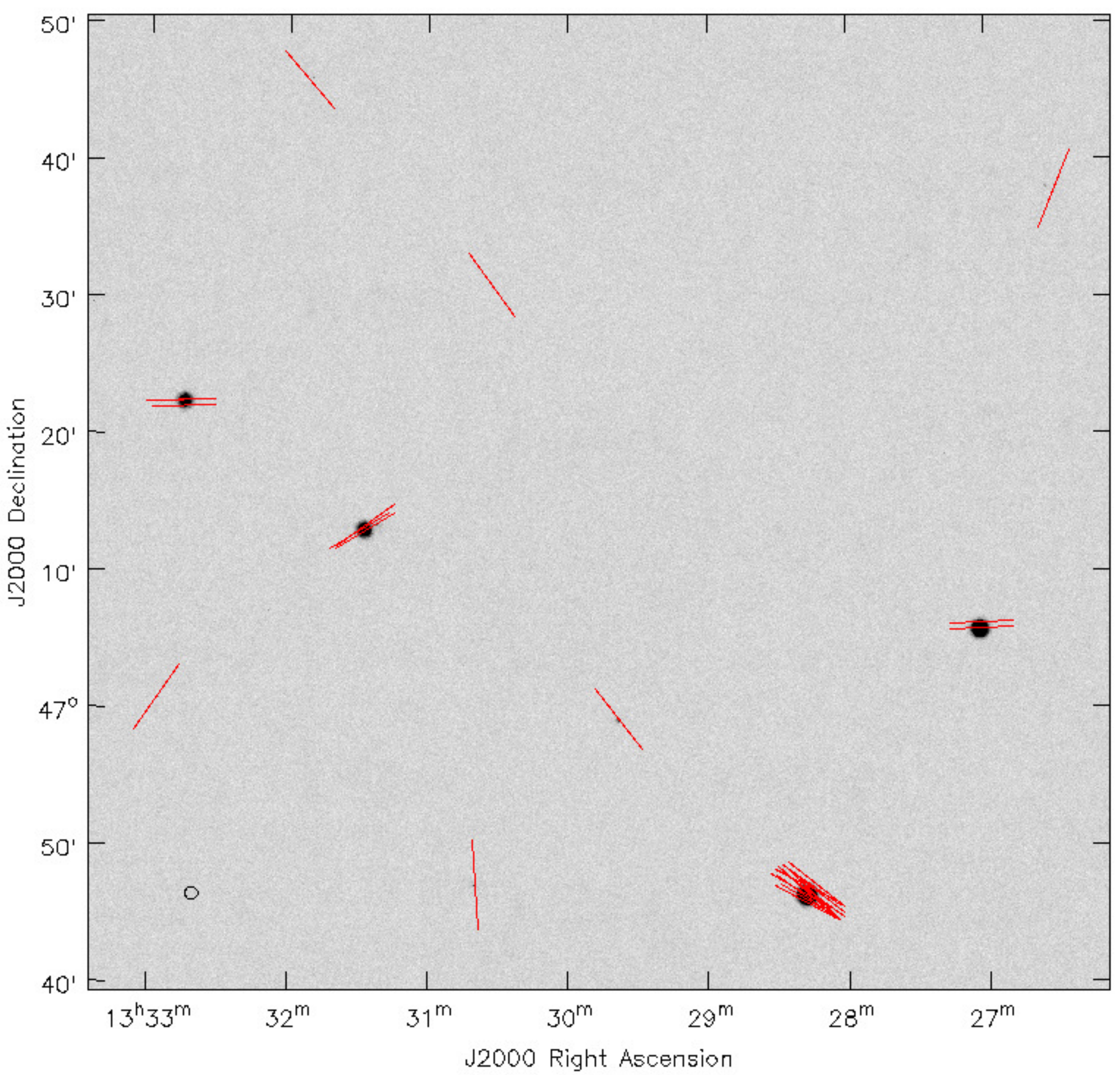}}
%  \vspace{-5pt} \\
  \caption[Polarisation vector orientation of the GMRT beam determined via RM Synthesis]{\small{Grayscale image of the bandwidth averaged polarised intensity, \(P = \sqrt{Q^2 + U^2}\), of the field surrounding M51 -- the image has not been corrected for the Stokes \(I\) beam. The centre of the image is also the phase-centre of the observation. Red vectors indicate the orientation of the direction-dependent instrumental polarisation. The vectors are those at a Faraday depth of \(0\)~rad~m\(^{-2}\), as obtained via RM Synthesis. The vectors are not proportional to polarised intensity -- variations in polarised intensity across the 610~MHz primary beam are shown in Figure \ref{chap6:PBefore}.}}
  \label{faradaydepth}
  \vspace{0pt}
\end{figure}

The application of RM Synthesis to the field of M51 can also be used to study the frequency-dependence of the direction-dependent polarisation. RM Synthesis is particularly useful as frequency-independent instrumental effects show up at a Faraday depth of \(0\)~rad~m\(^{-2}\) \citep{2005A&A...441.1217B}. The resulting output at a Faraday depth of \(0\)~rad~m\(^{-2}\) is shown in Figure \ref{faradaydepth}. The Faraday depth of each source is low, and no additional response was identified at high Faraday depths. The lack of direction-dependent instrumental polarisation at non-zero Faraday depths suggests that the wide-field polarisation beam has no significant frequency-dependence across an observing band at 610~MHz.

The RM Synthesis results from the field of M51 can also be used to study the geometry of the direction-dependent polarisation. The results of RM Synthesis suggest that the off-axis instrumental response is oriented in a predominantly radial manner, as shown in Figure \ref{faradaydepth}. It is not trivial to show that the vectors are truly oriented radially, as the observed EVPA's are calibrated relative to 3C286, and are also a combination of both physical and instrumental polarisation. For the more rigorous approach of holographic observations of the GMRT beam, see Farnes, (2012). Note that \citet{2009MNRAS.399..181P} show that this radial orientation is anticipated as leakage across the beam is expected to vary in a quadrupolar pattern. This is due to the GMRT's prime-focus feeds being mounted on four support legs which each have a width comparable to the observing wavelength. Consequently, unpolarised sources within the first null are expected to appear radially polarised, with sources beyond the null being tangentially polarised.

The radial orientation of the polarisation beam is useful for observations obtained during a full-track synthesis, as the radial polarisation will tend to average down over long integrations. This averaging down reduces the direction-dependent response within the half-power points at 610~MHz and assists considerably in reducing spurious polarisation.

\section{The Field of M51}
\label{chap5:M51}
\begin{figure*}
\centering
    \resizebox{180mm}{!}{\includegraphics[clip=false,trim=0cm 0cm 0cm 0cm]{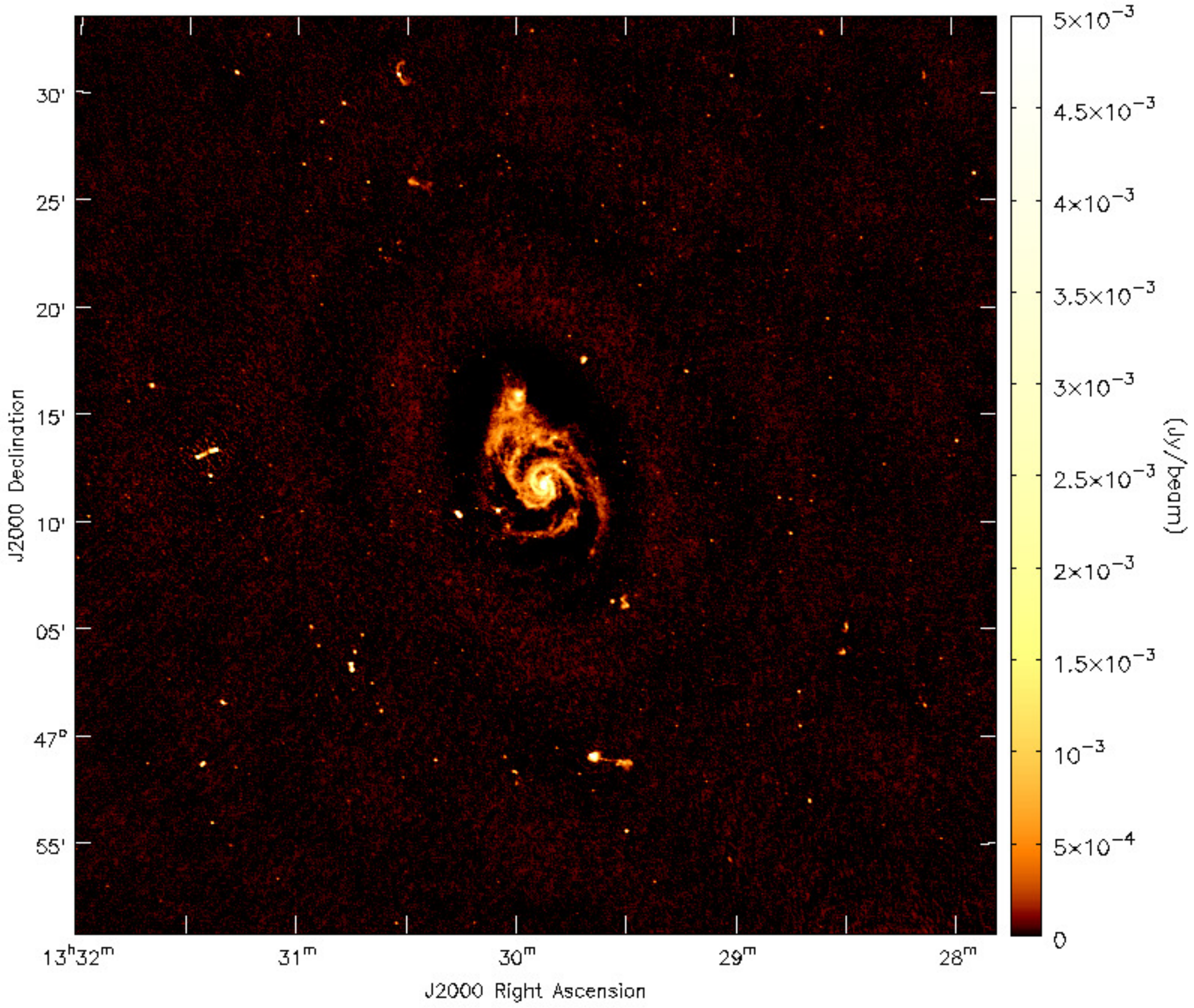}}
 % \vspace{-5pt} \\
  \caption[The galaxy M51 in Stokes I at 610~MHz]{\small{Stokes \(I\) image of the central portion of the field surrounding M51. The companion galaxy, NGC~5195, is clearly visible as the location of bright emission at the outer edge of the spiral arms to the N of M51. The pseudo-colour scale is in units of Jy~beam\(^{-1}\). The image has a resolution of \(5.7^{\prime\prime} \times 4.6^{\prime\prime}\) at a position angle of \(66^{\circ}\).}}
  \label{M51phasecentre}
  \vspace{0pt}
\end{figure*}

The galaxy M51 also provides the opportunity to assess the quality of polarisation calibration. The galaxy and its linear polarisation properties have been well-studied at a number of frequencies \citep[see e.g.][and references therein]{1992A&A...265..417H,2011MNRAS.412.2396F}. The galaxy has a well-defined spiral structure that is believed to be the result of interactions with the nearby companion galaxy, NGC~5195 \citep[e.g.][]{2010MNRAS.403..625D}. The total intensity data is shown in Figure \ref{M51phasecentre} at full resolution. The image has an rms noise of \(\sim23\)~\(\muup\)Jy~beam\(^{-1}\). 

\begin{figure*}[htb]
\centering
    \resizebox{180mm}{!}{\includegraphics[clip=false,trim=0cm 0cm 0cm 0cm]{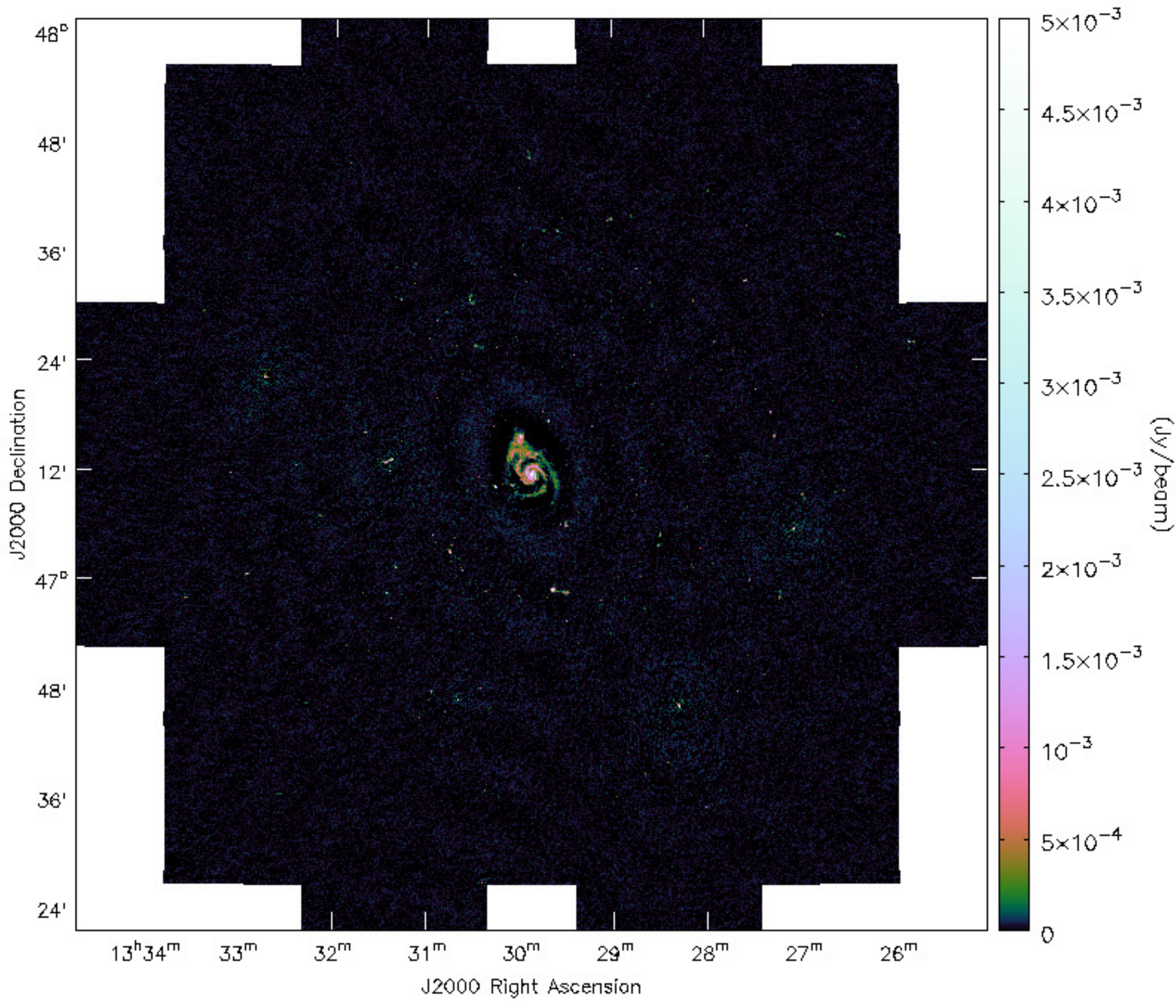}}
 % \vspace{-5pt} \\
  \caption[The field of view surrounding M51 in Stokes I]{\small{The wide-field Stokes \(I\) image of the field surrounding M51 at 610~MHz. The pseudo-colour scale uses the `cubehelix' colour scheme \citep{GreenREF}, and is in units of Jy~beam$^{-1}$. A correction for the effect of the Stokes \(I\) beam has not been applied. Also see Figure \ref{M51-FD0a} for the corresponding polarisation image.}}
  \label{M51-FD0}
  \vspace{0pt}
\end{figure*}

\begin{figure*}[htb]
\centering
    \resizebox{180mm}{!}{\includegraphics[clip=false,trim=0cm 0cm 0cm 0cm]{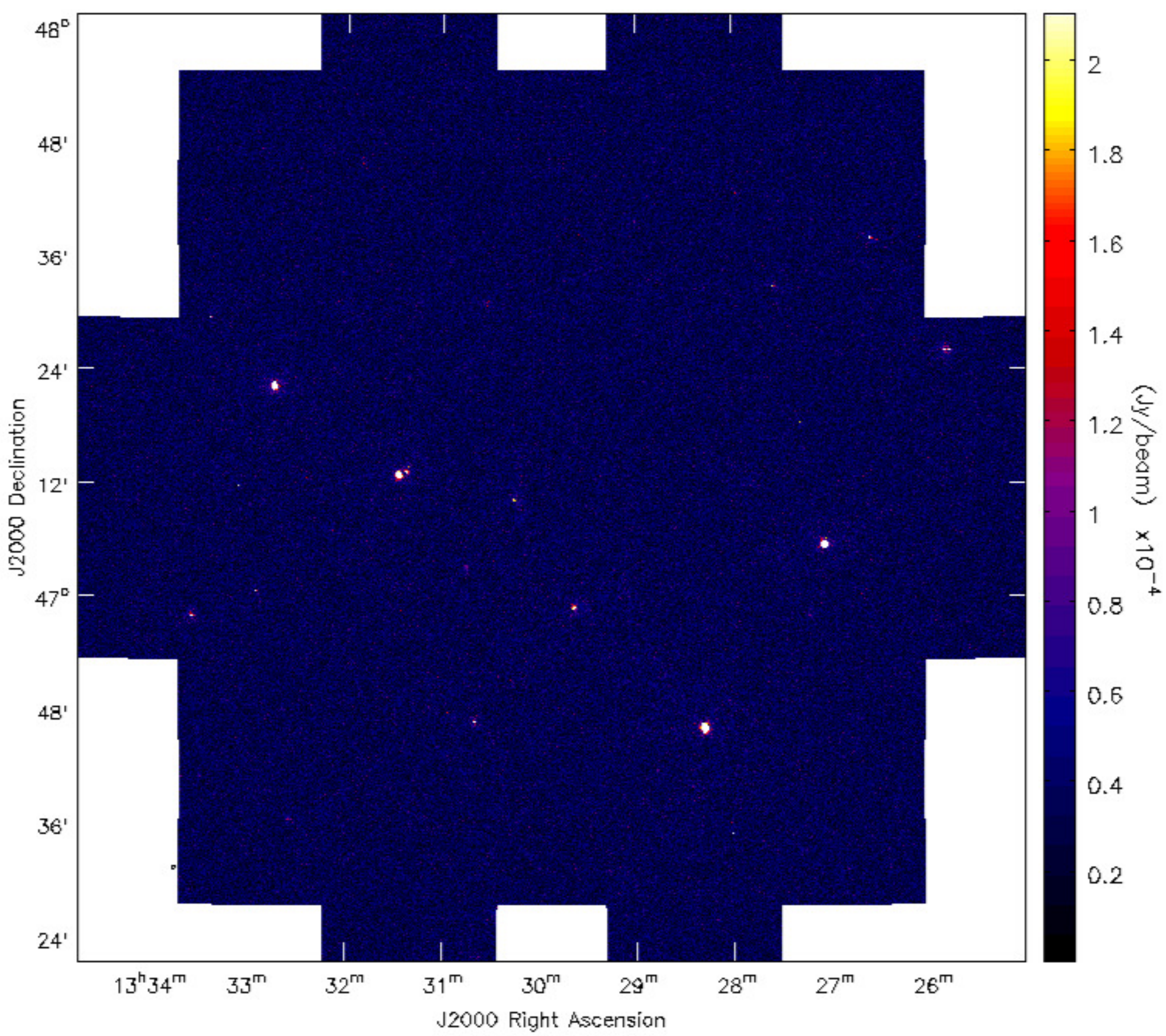}}
 % \vspace{-5pt} \\
  \caption[The field of view surrounding M51 at a low Faraday depth]{\small{The wide-field image of the field surrounding M51 at 610~MHz in polarised intensity for a Faraday depth of \(0\)~rad~m\(^{-2}\). The lack of instrumental artefacts, combined with the typically low Faraday depth of sources, and the large FWHM of the RMSF, mean that a Faraday depth of \(0\)~rad~m\(^{-2}\) is useful for displaying the polarised sources. A correction for the effects of the Stokes \(I\) primary beam has not been applied. Also see Figure \ref{M51-FD0} for the corresponding total intensity image.}}
  \label{M51-FD0a}
  \vspace{0pt}
\end{figure*}

Polarisation images were created using a \(uv\)-taper and natural weighting and have a resolution of \(24^{\prime\prime} \times 24^{\prime\prime}\). RM Synthesis was carried out on the entire field. A noise level of \(44\)~\(\muup\)Jy~beam\(^{-1}\)~rmsf\(^{-1}\) is achieved in the cleaned \(\phi\)-cubes. The FDs of the sources in the field were all found to be low (see Table 1). Due to the large FWHM of the RMSF, all of the polarised sources are visible at, or near, a Faraday depth of \(0\)~rad~m\(^{-2}\). The wide-field images in both Stokes \(I\), and at a Faraday depth of \(0\)~rad~m\(^{-2}\), are shown in Figures \ref{M51-FD0} and \ref{M51-FD0a}. Note the lack of instrumental artefacts from residual instrumental polarisation, and the substantial number of discrete polarised sources. 

At the available resolution, the galaxy M51 is completely depolarised down to the sensitivity limit at 610~MHz. The bright active nucleus is depolarised to a \(3\sigma\) upper limit of \(<0.14\)\%. The brighter spiral arms to the N and E of the galaxy are depolarised to a \(3\sigma\) upper limit of \(<1.1\)\%, while the fainter arms to the S and W are depolarised to \(<8.7\)\%. The bright emission detected in Stokes \(I\) from the companion galaxy NGC~5195 is depolarised to \(<0.4\)\%.

RM Synthesis was used to extract the polarised peak brightness and Faraday depth of all compact sources with a peak \(\ge8\sigma\) in the Faraday spectrum. The right ascension, declination, fractional polarisation, and peak Faraday depth of these sources are listed in Table \ref{tab:M51-FDs}. The listed co-ordinates are for the brightest pixel in Stokes \(I\). The polarisation fraction was only calculated for sources within the 10\% point of the Stokes \(I\) beam, i.e.\ at a radial distance \(\le35.6^{\prime}\) from the phase-centre. Residuals from the direction-dependent corrections likely cause an overestimation of \(\Pi\) for several of these sources. The calculated \(\Pi\) should be considered an upper limit, particularly for sources beyond the half-power points, i.e.\ at a radial distance \(\ge22.2^{\prime}\) from the phase-centre. Measurements of \(\Pi\) and the Faraday depth were calculated by fitting a Gaussian to the points immediately surrounding the peak in \(\phi\)-space. Previous Faraday rotation measurements from WSRT data at 18~cm and 22~cm \citep{2009A&A...503..409H} show reasonable agreement with the GMRT data.

\begin{table*}[!htb]
\caption[Details of compact polarised sources in the field of M51]{Positions in J2000, fractional polarisation, and Faraday depth of polarised sources in the field of M51. WSRT measurements are taken from \citet{2009A&A...503..409H}. All listed errors are the \(1\sigma\) uncertainties.} % title of Table
\centering % used for centering table
\begin{tabular}{c c c c c c} % centered columns (4 columns)
\hline\hline %inserts double horizontal lines
\# & RA. & DEC. & \(\Pi_{610}\) & FD\(_{\rm GMRT}\) & FD\(_{\rm WSRT}\) \\ [1ex] % inserts table
 &  &  & /\% & rad~m$^{-2}$ & rad~m$^{-2}$\\ [1ex] % inserts table
%heading
\hline % inserts single horizontal line
1 & 13h25m47.7s & \(47\)d26\(^{\prime}\)04.9\(^{\prime\prime}\) &  ---              & \(-16.97\pm0.03\)  & ---\\[1ex] 
2 & 13h27m03.3s & \(47\)d05\(^{\prime}\)43.3\(^{\prime\prime}\) &  \(6.32\pm0.14\)  & \(-8.11\pm0.07\)   & ---\\[1ex] 	  
3 & 13h27m57.2s & \(47\)d42\(^{\prime}\)51.1\(^{\prime\prime}\) &  ---              & \(16.83\pm0.13\)   & --- 	 \\[1ex] 
4 & 13h28m17.8s & \(46\)d46\(^{\prime}\)17.4\(^{\prime\prime}\) &  \(2.49\pm0.07\)  & \(-6.638\pm0.013\) & --- 	  \\[1ex]
5 & 13h29m39.4s & \(46\)d59\(^{\prime}\)10.2\(^{\prime\prime}\) &  \(1.72\pm0.09\)  & \(11.15\pm0.05\)   & \(14\pm1\) \\[1ex]
6 & 13h30m16.1s & \(47\)d10\(^{\prime}\)28.1\(^{\prime\prime}\) &  \(1.79\pm0.16\)  & \(33.52\pm0.03\)   & \(28\pm4\) \\[1ex]
7 & 13h30m32.4s & \(47\)d30\(^{\prime}\)53.1\(^{\prime\prime}\) &  \(6.9\pm0.5\)    & \(1.494\pm0.026\)  & --- 	  \\[1ex]
8 & 13h30m39.9s & \(46\)d47\(^{\prime}\)04.9\(^{\prime\prime}\) &  \(14.7\pm0.7\)   & \(-5.434\pm0.019\) & --- 	  \\[1ex]
9 & 13h30m44.9s & \(47\)d03\(^{\prime}\)08.3\(^{\prime\prime}\) &  \(1.98\pm0.20\)  & \(-3.24\pm0.04\)   & \(17\pm2\) \\[1ex]
10 & 13h30m45.2s & \(47\)d03\(^{\prime}\)25.1\(^{\prime\prime}\) &  \(2.74\pm0.24\)  & \(-15.97\pm0.03\)  & --- 	  \\[1ex]
11 & 13h31m22.4s & \(47\)d13\(^{\prime}\)22.5\(^{\prime\prime}\) &  \(6.84\pm0.17\)  & \(11.51\pm0.03\)   & \(9\pm1\)  \\[1ex]
12 & 13h31m27.4s & \(47\)d13\(^{\prime}\)00.6\(^{\prime\prime}\) &  \(7.22\pm0.25\)  & \(7.67\pm0.04\)    & \(3\pm1\)  \\[1ex]
13 & 13h32m45.1s & \(47\)d22\(^{\prime}\)22.6\(^{\prime\prime}\) &  \(6.83\pm0.24\)  & \(-1.41\pm0.03\)   & --- 	  \\[1ex]
14 & 13h33m24.9s & \(47\)d29\(^{\prime}\)35.3\(^{\prime\prime}\) &  ---              & \(1.71\pm0.28\)    & --- 	  \\[1ex]
15 & 13h33m35.2s & \(46\)d58\(^{\prime}\)07.5\(^{\prime\prime}\) &  ---              & \(0.93\pm0.23\)    & --- 	  \\[1ex]
% [1ex] adds vertical space
\hline %inserts single line
\end{tabular}
\label{tab:M51-FDs} % is used to refer this table in the text
\end{table*}

\section{Discussion and Conclusions}
\label{chap7:discussion}
We have shown the viability of the GMRT for full-polarisation, wide-field spectropolarimetry. This is possible despite the on-axis instrumental polarisation being of large amplitude and highly frequency-dependent at 610~MHz. Investigation of the nearby galaxy M51 has further shown that the polarisation calibration procedures work well. The galaxy itself and the nearby companion are both depolarised down to the sensitivity limit at 610~MHz. The bright active nucleus is depolarised to a \(3\sigma\) upper limit of \(<0.14\)\%. The brighter spiral arms to the N and E of the galaxy are depolarised to a \(3\sigma\) upper limit of \(<1.1\)\%, while the fainter arms to the S and W are depolarised to \(<8.7\)\%. The bright emission detected in Stokes \(I\) from the companion galaxy NGC~5195 is depolarised to \(<0.4\)\%. However, a number of bright polarised point sources are detected in linear polarisation across the field of view. The calculated Faraday depths of compact sources that have been detected in polarisation are in reasonable agreement with results obtained at higher frequencies \citep[see][]{2009A&A...503..409H}. Furthermore, five of these compact sources have not had their polarisation properties reported previously -- the Faraday depth and polarised fraction are reported here for the first time at 610~MHz.

The source located at 13h30m32.4s, 47\(^{\circ}\)30\(^{\prime}\)53.1\(^{\prime\prime}\) (\#7 in Table 1) is coincident with the galaxy cluster WHL~J133032.6+473053 located at a redshift of \(z=0.32\) \citep{2010ApJS..191..254H}. The detected source has a polarisation fraction of \(6.9\pm0.5\)\%, although the source of the polarised emission along the line of sight cannot be investigated in detail given the limited resolution. Furthermore, the source located at 13h32m45.1s, 47\(^{\circ}\)22\(^{\prime}\)22.6\(^{\prime\prime}\) (\#13 in Table 1) is coincident with the radio galaxy B3~1330+476 located at a redshift of \(z=0.67\) \citep{2005AJ....130..367S}. 

The GMRT has been used for both deep surveys at 610~MHz \citep[e.g.][]{2007MNRAS.376...1251G,2008MNRAS.383...75G,2008MNRAS.387.1037G,2009MNRAS.397...1101G,2010BASI.38...103G}  and large-area surveys such as the TIFR GMRT Sky Survey (TGSS) at 150~MHz\footnote{http://tgss.ncra.tifr.res.in/}. The TGSS is planned to detect up to two million sources in total intensity at 150~MHz. The resolution and sensitivity of the GMRT at low radio frequencies means that polarimetric radio surveys have the potential to be complementary to planned surveys with facilities such as ASKAP that explore a similar parameter space at higher frequencies and more Southern declinations \citep[e.g.][]{2008ExA....22..151J}. 

The full technical details of GMRT spectropolarimetry will be discussed in a future publication (Farnes et al.\ in preparation). Future observations with the polarisation mode of the GMRT could make useful predictions of the number density of polarised sources observable with the SKA and its pathfinders, and opens up the possibility to model magnetic fields and their evolution in populations of faint radio sources \citep[e.g.][]{2007mru..confE..69S,2009ASPC..407...12T}. Furthermore, it has been argued by \citet{2012A&A...543A.113B} that a combination of GMRT, LOFAR, and JVLA data will allow for magnetic structures at intermediate scales to be recognised -- comparable to the range of recognisable scales in Faraday space for SKA data. This would allow for much stronger constraints to be placed on the prevailing depolarisation mechanisms at low frequencies, allow for improved estimates of polarised source counts at low frequencies, and demonstrate the feasibility of using next-generation facilities, such as the SKA, for observing the RM-grids that will be essential for understanding cosmic magnetism \citep{2004NewAR..48.1003G}.

\section{Acknowledgements}
We thank the staff of the GMRT that made these observations possible. The GMRT is run by the National Centre for Radio Astrophysics of the Tata Institute of Fundamental Research.  J.S.F. acknowledges the support of the Australian Research Council through grant DP0986386. J.S.F. has been supported by the Science and Technology Facilities Council, UK.


\clearpage
\begin{thebibliography}
\bibitem[\protect\citeauthoryear{Anderson et al.}{2012}]{2012arXiv1203.2467A} Anderson, J., et al., 2012, ``The LOFAR Magnetism Key Science Project'', Proceedings of Magnetic Fields in the Universe: From Laboratory and Stars to Primordial Structures, 2011 August 21--27 in Zakopane/Poland. Edited by M.~Soida et al.

\bibitem[\protect\citeauthoryear{Beck et al.}{2012}]{2012A&A...543A.113B}
Beck, R., Frick, P., Stepanov, R., Sokoloff, D., 2012, A\&A, 543, A113.

\bibitem[\protect\citeauthoryear{Beck et al.}{2013}]{2013arXiv1302.0889B}
Beck, R., et al., 2013, ``The LOFAR View of Cosmic Magnetism'', Summary of a review talk given at the Annual Meeting of the Astronomische Gesellschaft "The Bright and the Dark Sides of the Universe", Hamburg, 2012 September 24--28.

 \bibitem[Brentjens(2008)Brentjens]{2008A&A...489...69B} 
Brentjens, M.~A., 2008, A\&A, 489, 69. 

\bibitem[\protect\citeauthoryear{Brentjens \& de Bruyn}{2005}]{2005A&A...441.1217B} 
Brentjens, M.~A., de Bruyn, A.~G., 2005, A\&A, 441, 1217.

% \bibitem[Burn(1966)Burn]{1966MNRAS.133...67B} 
%Burn, B.~J., 1966, MNRAS, 133, 67.
%
% \bibitem[Cotton(1993)Cotton]{1993AJ....106.1241C} 
%Cotton, W.~D., 1993, AJ, 106, 1241.
%
 \bibitem[Cotton(1994)Cotton]{CottonAIPSMemo} 
Cotton, W.~D., 1994, AIPS Memo \#86, ``Widefield Polarisation Correction of VLA Snapshot Images at 1.4~GHz''.

 \bibitem[Dobbs et al.(2010)Dobbs et al.]{2010MNRAS.403..625D}
Dobbs, C.~L., Theis, C., Pringle, J.~E., Bate, M.~R., 2010, MNRAS, 403, 625.

 \bibitem[Farnes(2012)Farnes]{mythesis} 
Farnes, J.~S., 2012, ``Polarimetric Observations at Low Radio Frequencies'', PhD thesis, University of Cambridge.

 \bibitem[Farnsworth et al.(2011)Farnsworth, Rudnick, \& Brown]{2011AJ....141..191F} 
Farnsworth, D., Rudnick, L., Brown, S., 2011, AJ, 141, 191. 

 \bibitem[Fletcher et al.(2011)Fletcher et al.]{2011MNRAS.412.2396F} 
Fletcher, A., et al., 2011, MNRAS, 412, 2396. 

 \bibitem[Frick et al.(2010)Frick et al.]{2010MNRAS.401L..24F} 
Frick, P., Sokoloff, D., Stepanov, R., Beck, R., 2010, MNRAS, 401, L24.

 \bibitem[Gaensler et al.(2004)Gaensler, Beck, \& Feretti]{2004NewAR..48.1003G} 
Gaensler, B.~M., Beck, R., Feretti, L., 2004, NewAR, 48, 1003. 

 \bibitem[Garn et al.(2007)Garn et al.]{2007MNRAS.376...1251G}
Garn, T.~S., et al., 2007, MNRAS, 376, 1251.

 \bibitem[Garn et al.(2008a)Garn et al.]{2008MNRAS.383...75G}
Garn, T.~S., Green, D.~A., Riley, J.~M., Alexander, P., 2008, MNRAS, 383, 75.

 \bibitem[Garn et al.(2008b)Garn et al.]{2008MNRAS.387.1037G}
Garn, T.~S., Green, D.~A., Riley, J.~M., Alexander, P., 2008, MNRAS, 387, 1037.

% \bibitem[Garn(2009)Garn]{2009PhDT.........3G} 
%Garn, T.~S., 2009, ``610~MHz observations of galaxy evolution'', PhD thesis, University of Cambridge.
%
 \bibitem[Garn et al.(2009)Garn et al.]{2009MNRAS.397...1101G}
Garn, T.~S., Green, D.~A., Riley, J.~M., Alexander, P., 2009, MNRAS, 397, 1101.

 \bibitem[Garn et al.(2010)Garn et al.]{2010BASI.38...103G}
Garn, T.~S., Green, D.~A., Riley, J.~M., Alexander, P., 2010, BASI, 38, 103.

 \bibitem[George et al.(2012)George, Stil, \& Keller]{2011arXiv1106.5362G} 
George, S.~J., Stil, J.~M., Keller, B.~W., 2012, PASA, 29, 214. 

%\bibitem[\protect\citeauthoryear{Gopal-Krishna et al.}{2012}]{2012MNRAS.423.1053G} Gopal-Krishna, et al., 2012, MNRAS, 423, 1053 
%
 \bibitem[Green(2011)Green]{GreenREF} 
Green, D.~A., 2011, BASI, 39, 289.

 \bibitem[Gupta(2011)Gupta]{GuptaREF}
Gupta,~Y., ``Observatory report for the GMRT'', General Assembly and Scientific Symposium, 2011 XXXth URSI, 13--20 August, 2011.

 \bibitem[Hamaker et al.(1996)Hamaker et al.]{HamakerREF} 
Hamaker, J.~P., Bregman, J.~D., Sault, R.~J., 1996, A\&AS, 117, 137.

 \bibitem[Hao et al.(2010)Hao et al.]{2010ApJS..191..254H} 
Hao, J., et al., 2010, ApJS, 191, 254. 

 \bibitem[Heald et al.(2009)Heald, Braun, \& Edmonds]{2009A&A...503..409H} 
Heald, G., Braun, R., Edmonds, R., 2009, A\&A, 503, 409. 

 \bibitem[Heiles(2002)Heiles]{2002ASPC.proc..278S} 
Heiles, C., 2002, ``A Heuristic Introduction to Radioastronomical Polarization'', Single-Dish Radio Astronomy: Techniques and Applications, ASP Conference Proceedings, Vol.~278. Edited by Snezana Stanimirovic, Daniel Altschuler, Paul Goldsmith, and Chris Salter. San Francisco: Astronomical Society of the Pacific, 2002, p.131--152.

 \bibitem[Heiles et al.(2003)Heiles et al.]{HeilesGBTMemo} 
Heiles, C., Robishaw, T., Troland, T., Anish Roshi, D., 2003, ``GBT Commissioning Memo \#23''.

 \bibitem[Horellou et al.(1992)Horellou et al.]{1992A&A...265..417H} 
Horellou, C., et al., 1992, A\&A, 265, 417.

\bibitem[\protect\citeauthoryear{Johnston et al.}{2008}]{2008ExA....22..151J} 
Johnston, S., et al., 2008, ExA, 22, 151.

\bibitem[\protect\citeauthoryear{Joshi \& Chengalur}{2010}]{joshipaper} 
Joshi, S., \& Chengalur, J.~N., 2010, ``Polarimetric images with the GMRT'', Proceedings of the ISKAF2010 Science Meeting, June 10--14, 2010. Assen, the Netherlands. Published online at SISSA, Proceedings of Science, p.30.

 \bibitem[Pen et al.(2009)Pen et al.]{2009MNRAS.399..181P} 
Pen, U.-L., et al., 2009, MNRAS, 399, 181. 

\bibitem[\protect\citeauthoryear{Perley et al.}{2009}]{2009IEEEP..97.1448P} 
Perley, R., et al., 2009, IEEEP, 97, 1448.

 \bibitem[Schneider et al.(2005)Schneider et al.]{2005AJ....130..367S} 
Schneider, D.~P., et al., 2005, AJ, 130, 367. 

 \bibitem[Simmons \& Stewart(1985)Simmons \& Stewart]{1985A&A...142..100S} 
Simmons, J.~F.~L., Stewart, B.~G., 1985, A\&A, 142, 100. 

 \bibitem[Stil et al.(2007)Stil et al.,]{2007mru..confE..69S} 
Stil, J.~M., Taylor, A.~R., Krause, M., Beck, R., 2007, ``Polarisation of mJy radio sources'', From Planets to Dark Energy: The Modern Radio Universe. October 1--5 2007, The University of Manchester, UK. Published online at SISSA, Proceedings of Science, p.69.

 \bibitem[Taylor(2009)Taylor]{2009ASPC..407...12T}
Taylor, A.~R., 2009, ``High-Resolution Spectro-Polarimetric Radio Surveys'', The Low-Frequency Radio Universe, ASP Conference Series, Vol.~407. Proceedings of the conference, held at National Centre for Radio Astrophysics, TIFR, Pune, India, December 8--12, 2008. Edited by D.~J.~Saikia, D.~A.~Green, Y.~Gupta, and T.~Venturi. San Francisco: Astronomical Society of the Pacific, 2009, p.12.

 \bibitem[Taylor et al.(2010)Taylor \& Salter]{2010ASPC..438..402T}
Taylor, A.~R., Salter, C.~J., ``GALFACTS: The G-ALFA Continuum Transit Survey'', The Dynamic Interstellar Medium: A Celebration of the Canadian Galactic Plane Survey. Proceedings of the conference, held at the Naramata Centre, Naramata, British Columbia, Canada, June 6--10, 2010. Edited by R.~Kothes, T.~L.~Landecker, and A.~G.~Willis. San Francisco: Astronomical Society of the Pacific, 2010, p.402.

 \bibitem[Tinbergen(1996)Tinbergen]{1996aspo.book.....T} 
Tinbergen, J., 1996, ``Astronomical Polarimetry'', Cambridge University Press.

\end{thebibliography}
\end{document}